\documentclass[12pt]{iopart}

\usepackage{iopams}  
\usepackage{setstack}
\usepackage{subfigure}
\usepackage{subeqn}
\usepackage{graphicx}
\graphicspath{{./}{./images/}{./images_all/}{./images_all/interface_coupled/}}

\begin{document}

\title[]{\Large{Fast Track Communication}\\\vspace*{5mm}Electric-field induced magnetization switching in interface-coupled multiferroic heterostructures: A highly-dense, non-volatile, and ultra-low-energy computing paradigm}

\author{Kuntal Roy}
\address{School of Electrical and Computer Engineering, Purdue University, West Lafayette, IN 47907, USA.\\**Some works were performed prior joining at Purdue University.}

\ead{royk@purdue.edu}

\begin{abstract}
Electric-field induced magnetization switching in multiferroic magnetoelectric devices is promising for beyond Moore's law computing. We show here that interface-coupled multiferroic heterostructures, i.e., a ferroelectric layer coupled with a ferromagnetic layer, are particularly suitable for highly-dense, non-volatile, and ultra-low-energy computing. By solving stochastic Landau-Lifshitz-Gilbert equation of magnetization dynamics in the presence of room-temperature thermal fluctuations, we demonstrate that error-resilient switching of magnetization is possible in sub-nanosecond delay while expending a minuscule amount of energy of $\sim$1 attojoule. Such devices can be operated by drawing energy from the environment without the need for an external battery.
\end{abstract}

\maketitle

\section{Introduction}
\label{sec:introduction}

Electric field induced magnetization switching mechanism holds profound promise for computing in beyond Moore's law era~\cite{roy13_spin,roy11}. In multiferroic magnetoelectrics, application of an electric field can rotate the magnetization via converse magnetoelectric effect, however, such materials in \emph{single-phase} usually have issues of weak coupling between polarization and magnetization and they operate only at low temperatures~\cite{RefWorks:558,RefWorks:665}. Although, new concepts may come along on switching in single-phase materials~\cite{RefWorks:667,RefWorks:669}, strain-mediated multiferroic heterostructures~\cite{roy11,RefWorks:558,Refworks:164,Refworks:165} consisting of a piezoelectric layer coupled to a magnetostrictive nanomagnet, is shown to be very effective. With appropriate choice of materials, when a voltage of few millivolts is applied across such heterostructures, the piezoelectric layer gets strained and the strain is elastically transferred to the magnetostrictive layer rotating its magnetization. Such switching mechanism dissipates a minuscule amount of energy of $\sim$1 attojoule (aJ) in sub-nanosecond switching delay at room-temperature~\cite{roy11_6}. This study has opened up a new field called \emph{straintronics}~\cite{roy13_spin,roy13,roy14} and experimental efforts to demonstrating electric-field induced magnetization switching are considerably emerging~\cite{RefWorks:559,RefWorks:551,RefWorks:609,RefWorks:611}. In a strain-mediated multiferroic heterostructure, the strain transferred by the piezoelectric layer to the magnetostrictive nanomagnet, can only rotate the magnetization $90^\circ$ in \emph{steady-state} consideration~\cite{roy13_2}. Although there are proposals of $90^\circ$ switching mechanism~\cite{RefWorks:520,RefWorks:551,RefWorks:559}, it is explained that a complete $180^\circ$ switching is possible if we consider the \emph{dynamics} of magnetization into account rather than \emph{assuming} steady-state scenario; basically the magnetization's excursion out of magnet's plane provides an equivalent asymmetry to cause switching in the correct direction~\cite{roy11,roy13_2}.

\begin{figure}
\centering
\includegraphics[width=80mm]{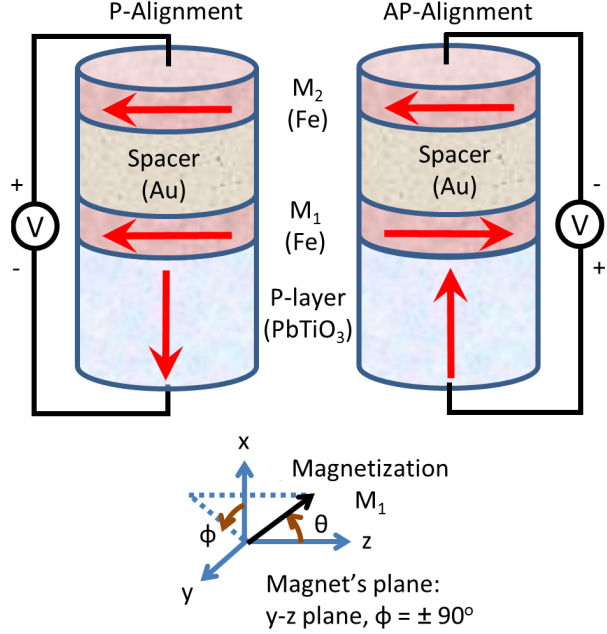}
\caption{\label{fig:schematic_interface_coupled} Schematics of the interface-coupled multiferroic magnetoelectric devices (see Ref.~\cite{RefWorks:649}). The unique coupling between the polarization in the P-layer and the trilayer $M_1$/spacer/$M_2$ allows the polarization direction to dictate the magnetic ground state in the trilayer. If the polarization points downward ($P_\downarrow$), P-alignment in the trilayer is preferred while an upward polarization ($P_\uparrow$) prefers the AP-alignment. Application of a voltage with correct polarity can switch the polarization and hence the magnetization $M_1$ gets switched too due to interface coupling. At the bottom of the figure, the axis assignment for the dynamical motion of magnetization $M_1$ in standard spherical coordinate system is shown.} 
\end{figure}

Although the aforesaid \emph{strain-mediated} mechanism in multiferroic heterostructures is promising, it would be of substantial interest if there exists a strong coupling between the polarization and magnetization at the heterostructure interface. Recently, \emph{interface-coupled} multiferroics are proposed based on density functional theory (DFT) of first-principles calculations~\cite{RefWorks:649}. Despite lacking experimental verification for this specific case as of now, considering that first-principles calculations have been proved to be very useful~\cite{RefWorks:512} and with the experimental progress on similar front~\cite{RefWorks:793,RefWorks:404,RefWorks:795,RefWorks:796,RefWorks:797} (also using ferromagnetic oxides~\cite{RefWorks:792,RefWorks:798} rather than ferromagnetic metals), there is a considerable interest on such coupling mechanism~\cite{RefWorks:649,RefWorks:676,RefWorks:680,RefWorks:681,RefWorks:682,RefWorks:683}. Figure~\ref{fig:schematic_interface_coupled} depicts the interface coupling between polarization and magnetization in a multiferroic heterostructure. The polarization direction in the P-layer determines the ground state of the trilayer $M_1$/spacer/$M_2$. For polarization P$_\downarrow$, parallel alignment (P-alignment) in the trilayer is preferred, while for the polarization P$_\uparrow$, antiparallel alignment (AP-alignment) in the trilayer is preferred. This unique coupling phenomenon alongwith electric-field induced polarization switching makes the switching of magnetization in the $M_1$-layer non-volatile. Also, if a voltage with certain polarity is applied and maintained, the state of the system remains unaltered. This is advantageous over the strain-mediated switching, which just toggles the magnetization states. There are other exchanged coupled systems with an insulating spacer layer, but the interlayer exchange coupling energy is small~\cite{RefWorks:725,RefWorks:726}. Also, there are other schemes with non-magnetic spacer layer to preserve large interlayer exchange coupling, however, the electric field required is high and also the switching is volatile~\cite{RefWorks:727}.

Here, we study the magnetization dynamics in the interface-coupled multiferroic heterostructures by solving stochastic Landau-Lifshitz-Gilbert equation in the presence of room-temperature thermal fluctuations. Such phenomenological study of switching has been very useful to understand the performance metrics of magnetic devices~\cite{roy11,roy11_6,roy13_spin}. First, we model the interfacial anisotropy in the interface-coupled multiferroic heterostructures and then we analyze the dynamics of magnetization, which shows that switching in sub-nanosecond delay is possible while expending only $\sim$1 aJ of energy at room-temperature. The strong interface anisotropy makes the switching error-resilient and fast, and it allows us to work with nanomagnets with small dimensions, i.e., the magnetization is stable with $\sim$10 nm lateral dimensions even in the presence of room-temperature thermal fluctuations. Such superior performance metrics of area, delay, and energy are particularly suitable for computing in beyond Moore's law era.

\section{Model}
\label{sec:model}

Figure~\ref{fig:schematic_interface_coupled} shows the schematics of the interface-coupled multiferroic heterostructure devices and the axis assignment for the orientation of magnetization $M_1$. The standard spherical coordinate system with $\theta$ as polar angle and $\phi$ as azimuthal angle is utilized. The magnetization $M_1$ orients along $\theta=180^\circ$ if polarization points downward ($P_\downarrow$), while $M_1$ is along $\theta=0^\circ$ if polarization points upward ($P_\uparrow$). The elliptical cross-section of $M_1$ lies on the $y$-$z$ plane ($\phi=\pm90^\circ$) with its major axis pointing to $z$-direction and minor axis in $y$-direction. The dimensions of the nanomagnet $M_1$ along the $z$-,$y$-, and $x$-axis are $a$, $b$, and $l$, respectively. So the nanomagnet's volume $\Omega=(\pi/4)abl$. Magnetization's ground states reside on the magnet's plane ($y$-$z$ plane, $\phi=\pm90^\circ$), however, during the \emph{dynamical motion}, magnetization can deflect out of magnet's plane and any deflection from $\phi=\pm90^\circ$ is termed as out-of-plane excursion.

The interface anisotropy energy in the nanomagnet $M_1$ per unit volume is modeled as
\begin{equation}
E_{I}(\theta,t) = - M H_{I}(t)\,cos\theta,
\label{eq:anisotropy_interface}
\end{equation}
where $M=\mu_0 M_s$, $M_s$ is the saturation magnetization, and $H_I$ is the interfacial anisotropy field. If $H_I=-H_{I,max}$, the ground state of magnetization $M_1$ points along $\theta=180^\circ$ and if we vary $H_I$ from $-H_{I,max}$ to $H_{I,max}$, the ground state orients along $\theta=0^\circ$. The total anisotropy of the magnet is the sum of the interface anisotropy alongwith the other anisotropies like magnetocrystalline anisotropy and shape anisotropy~\cite{roy11,roy11_6}, however, due to strong interfacial anisotropy compared to the other anisotropies, we consider only the interfacial anisotropy (i.e., $E_{total} \simeq E_{I}$) for brevity. We assume that the nanomagnet has a shape of an elliptical cylinder with the ellipse's major axis along the $z$-direction, so that the magnetic easy axis becomes along the $\pm z$-direction.

The magnetization $M$ of the single-domain nanomagnet $M_1$ (having constant magnitude of magnetization but a variable direction) can be represented by the unit vector in the radial direction $\mathbf{\hat{e}_r}$ in spherical coordinate system ($r$,$\theta$,$\phi$), i.e., $\mathbf{n_m} =\mathbf{M}/|\mathbf{M}| = \mathbf{\hat{e}_r}$. The other two unit vectors in the spherical coordinate system are $\mathbf{\hat{e}_\theta}$ and $\mathbf{\hat{e}_\phi}$ for $\theta$- and $\phi$-rotations, respectively. The torque $\mathbf{T_I}$ acting on the magnetization due to interface anisotropy can be derived from the gradient of the energy and is given by

\begin{equation}
\mathbf{T_{I}}(\theta,t) = - \mathbf{n_m} \times \nabla E_I (\theta,t) = - M H_{I}(t)\,sin\theta \, \mathbf{\hat{e}_\phi}. 
\label{eq:torque_interface}
\end{equation}
Note that the torque $\mathbf{T_{I}}$ acts along the out-of-plane direction, so that the magnetization can delflect out of magnet's plane (i.e., $\phi$ can deflect from $\pm90^\circ$). 

The effect of random thermal fluctuations is incorporated via a random magnetic field $\mathbf{h}(t)= h_x(t)\mathbf{\hat{e}_x} + h_y(t)\mathbf{\hat{e}_y} + h_z(t)\mathbf{\hat{e}_z}$, where $h_i(t)$ ($i=x,y,z$) are the three components of the random thermal field in Cartesian coordinates. We assume the properties of the random field $\mathbf{h}(t)$ as described in Ref.~\cite{RefWorks:186}. The random thermal field can be written as~\cite{RefWorks:186}
\begin{equation}
h_i(t) = \sqrt{\frac{2 \alpha kT}{|\gamma| M \Omega \Delta t}} \; G_{(0,1)}(t) \quad (i \in x,y,z),
\label{eq:ht}
\end{equation}
\noindent
where $\alpha$ is the dimensionless phenomenological Gilbert damping parameter, $\gamma$ is the gyromagnetic ratio for electrons, $1/\Delta t$ is the attempt frequency of thermal noise, $\Omega$ is the volume, and the quantity $G_{(0,1)}(t)$ is a Gaussian distribution with zero mean and unit variance. 

The thermal field and the corresponding torque acting on the magnetization per unit volume can be written as $\mathbf{H_{TH}}(\theta,\phi,t)=P_\theta(\theta,\phi,t)\,\mathbf{\hat{e}_\theta}+P_\phi(\theta,\phi,t)\,\mathbf{\hat{e}_\phi}$ and $\mathbf{T_{TH}}(\theta,\phi,t)=\mathbf{n_m} \times \mathbf{H_{TH}}(\theta,\phi,t)$, respectively, 
where
\begin{eqnarray}
P_\theta(\theta,\phi,t) &=& M  \lbrack h_x(t)\,cos\theta\,cos\phi + h_y(t)\,cos\theta sin\phi - h_z(t)\,sin\theta \rbrack,\\
P_\phi(\theta,\phi,t) &=& M  \lbrack h_y(t)\,cos\phi -h_x(t)\,sin\phi\rbrack.
\label{eq:thermal_parts}
\end{eqnarray}

The magnetization dynamics under the action of the torques $\mathbf{T_{I}}$ and 
$\mathbf{T_{TH}}$ is described by the stochastic Landau-Lifshitz-Gilbert (LLG) equation as follows.
\begin{equation}
\frac{d\mathbf{n_m}}{dt} - \alpha \left(\mathbf{n_m} \times \frac{d\mathbf{n_m}}{dt} \right)
 = -\frac{|\gamma|}{M} \left\lbrack \mathbf{T_I} +  \mathbf{T_{TH}}\right\rbrack.
\end{equation}

Solving the above equation analytically, we get the following coupled equations of magnetization dynamics for $\theta$ and $\phi$:
\begin{equation}
\left(1+\alpha^2 \right) \frac{d\theta}{dt} = \frac{|\gamma|}{M} \, \lbrack  - \alpha M H_{I}(t)\, sin\theta 
				+ \left(\alpha P_\theta(\theta,\phi,t) + P_\phi (\theta,\phi,t) \right) \rbrack,
 \label{eq:theta_dynamics}
\end{equation}
\begin{eqnarray}
\left(1+\alpha^2 \right) \frac{d \phi}{dt} &=& \frac{|\gamma|}{M} \, \lbrack M H_{I}(t)
					- {[sin\theta]^{-1}} 
					\left(P_\theta (\theta,\phi,t) - \alpha P_\phi (\theta,\phi,t) \right) \rbrack  \nonumber \\
	&& \hspace*{7cm} (sin\theta \neq 0).
\label{eq:phi_dynamics}
\end{eqnarray}
We solve the above two coupled equations numerically to track the trajectory of magnetization over time, in the presence of room-temperature thermal fluctuations. 

From Eqs.~(\ref{eq:theta_dynamics}) and~(\ref{eq:phi_dynamics}), we see that the torque acting in the $\phi$-direction is much more higher than the torque exerted in the $\theta$-direction since the damping parameter $\alpha \ll 1$. Although, the nanomagnet has small dimension along its thickness (i.e., $l \ll b < a$ and the demagnetization factors $N_{dx} \gg N_{dy} > N_{dz}$), magnetization cannot remain on the magnet's plane ($y$-$z$ plane, $\phi=\pm90^\circ$) since the interface coupling energy is a few orders of magnitude higher than the shape anisotropy energy. Thus, the magnetization keeps rotating in the $\phi$-direction, but it also traverses towards the anti-parallel direction in $\theta$-space (e.g., $\theta \simeq 180^\circ$ to $\theta \simeq 0^\circ$) due to damping [see Eq.~(\ref{eq:theta_dynamics})].

Note that exactly at $\theta=180^\circ$ or $0^\circ$, the torque acting on the magnetization due to interface anisotropy [Eq.~(\ref{eq:torque_interface})] is exactly zero, however, thermal fluctuations can scuttle the magnetization from these points to initiate switching. At the very start of switching, the initial orientation of magnetization is not a fixed value rather a distribution due to thermal agitations. Such distribution is considered during simulations. Hence, thermal fluctuations affect the switching when magnetization starts to switch as well as during the course of switching.

The energy dissipated in the nanomagnet due to Gilbert damping can be expressed as  $E_d = \int_0^{\tau}P_d(t) dt$, where $\tau$ is the switching delay and $P_d(t)$ is the power dissipated at time $t$ per unit volume given by
\begin{equation}
P_d(t) = \frac{\alpha \, |\gamma|}{(1+\alpha^2) M} \, |\mathbf{T_I} (\theta(t), t)|^2 .
\label{eq:power_dissipation}
\end{equation}
Thermal field with mean zero does not cause any net energy dissipation but it causes variability in the energy dissipation by scuttling the trajectory of magnetization.

\section{Results}
\label{sec:results}

As depicted in the Fig.~\ref{fig:schematic_interface_coupled}, the nanomagnets ($M_1$ and $M_2$ layers) are made of Fe, while the ferroelectric $P$-layer is made of $PbTiO_3$~\cite{RefWorks:649,RefWorks:676,RefWorks:677,RefWorks:672}. The spacer layer is made of $Au$ and the thicknesses of the trilayer $M_1$/spacer/$M_2$ are 1/4/1 monolayers~\cite{RefWorks:649}. The Fe layer has a unit cell length of 0.287 nm and it has the following material parameters: saturation magnetization ($M_s$) = 1e6 A/m, and damping parameter ($\alpha$) = 0.01~\cite{RefWorks:650,RefWorks:674,RefWorks:675}. The elliptical lateral cross-section has a dimension of $15\,nm \times 7\,nm$. The $P$-layer has a unit cell length of 0.388 nm and it has 5 layers in the vertical direction ($x$-direction, Fig.~\ref{fig:schematic_interface_coupled})~\cite{RefWorks:649} while the lateral dimensions are same as of the nanomagnets. The energy difference between the P-alignment and AP-alignment is 10 mev/atom~\cite{RefWorks:649}, and the absolute value of energy turns out to be about 10 ev or 385 kT at room-temperature. This huge interface coupling makes potential landscape of $M_1$ \emph{monostable} at $\theta=180^\circ$ or at $\theta=0^\circ$ depending on the $P$-alignment or the $AP$-alignment in the trilayer, respectively. Hence, no spontaneous switching of magnetization between $\theta=180^\circ$ and $\theta=0^\circ$ can take place. The shape anisotropic energy barrier is a few orders of magnitude lower than this interface coupling energy and thus consideration of shape anisotropy does not make any significant difference, however, it is included during the simulations.

\begin{figure}
\centering
\includegraphics[width=80mm]{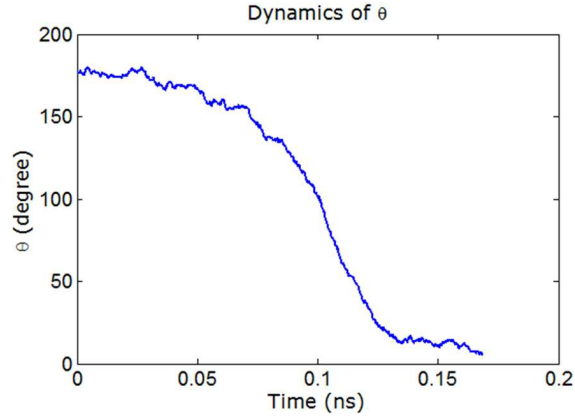}
\caption{\label{fig:thermal_single_run} A sample dynamics of magnetization while switching from $\theta \simeq 180^\circ$ to $\theta \simeq 0^\circ$ in the presence of room-temperature (300 K) thermal fluctuations. The ramp period is 100 ps and the switching delay is 168.5 ps. The energy dissipation due to Gilbert damping is 1.42 aJ.} 
\end{figure}

Modeling the $P$-layer as a parallel plate capacitor and using a relative dielectric constant of 1000~\cite{RefWorks:673}, the capacitance $C$ of the layer becomes $\sim$0.4 fF. If the $P$-layer is accessed with a 10 $\mu m$ long silver wire with resistivity $\sim$2.6 $\mu \Omega$-$cm$~\cite{interconnect}, the resistance $R$ becomes $\sim$3 $k\Omega$. Therefore, the $RC$ time constant is of the order of 1 ps. The ferroelectric $PbTiO_3$ has a coercive voltage of 20 MV/m~\cite{RefWorks:670} and hence a voltage $V\simeq40$ mv is required to switch the polarization. (Note that the voltage required to switch the traditional charge based devices is of the order of 1 V~\cite{RefWorks:729}.) Polarization switching is possible in less than 100 ps~\cite{RefWorks:430} and a voltage ramp with period $T = 100$ ps or more is considered to enforce the quasistatic (adiabatic) assumption ($T \gg RC$). Without any adiabatic assumption, the metric $CV^2$ is $0.5$ aJ and hence the energy dissipation in the external circuitry that applies the voltage is miniscule. With 100 ps ramp period, the ``$CV^2$" dissipation becomes a negligible value of 0.01 aJ~\cite{roy11_2,roy11_6}. Note that we do not calculate any \emph{standby} leakage current through the thin ferroelectric since the device operation is non-volatile, i.e., it is possible to turn off the voltage without loosing the information. Also, during the active mode of operation, the tunneling current is small ($<$ 1 nA~\cite{RefWorks:792}), leading to negligible energy dissipation. Note that ferroelectric fatigue may make the coercive field higher~\cite{RefWorks:800,RefWorks:799}, however, since the energy dissipation due to applied voltage is miniscule, it does not appear to be a bottleneck provided the polarization switches and the interface coupling between the $P$-layer and the trilayer still persists.

Figure~\ref{fig:thermal_single_run} shows a sample dynamics of magnetization in the presence of room-temperature (300 K) thermal fluctuations. The ramp period is considered to be 100 ps and the switching has completed in less than 175 ps. During the course of switching magnetization has temporarily backtracked due to random thermal kicks, however, the strong interface anisotropy has enforced magnetization to switch from $\theta \simeq 180^\circ$ to $\theta \simeq 0^\circ$.

\begin{figure}
\centering
\includegraphics[width=80mm]{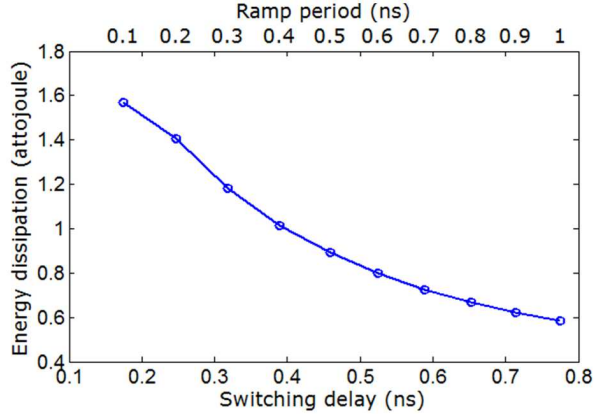}
\caption{\label{fig:delay_energy_ramp} Switching delay-energy trade-off as a function of ramp period (upper axis). For a faster ramp, the switching becomes faster but the energy dissipation goes higher. Each point is generated from 10000 simulations in the presence of room-temperature (300 K) thermal fluctuations and the average values of switching delays and energy dissipations are plotted. For 0.1 ns ramp period, the average (max) switching delay is 175.3 ps (330.8 ps), while the mean energy dissipation is 1.56 aJ. For a slower ramp with period 1 ns, the average (max) switching delay is 775.2 ps (1003.5 ps), while the mean energy dissipation is 0.58 aJ.} 
\end{figure}

Figure~\ref{fig:delay_energy_ramp} plots the average switching delay versus average energy dissipation for different ramp periods (0.1 ns -- 1 ns). A moderately large number of simulations (10000) in the presence of room-temperature (300 K) thermal fluctuations are performed to generate each point in the curve. When the magnetization reaches $\theta \leq 5^\circ$, the switching is deemed to have completed. Note that as we vary the ramp period of applied voltage across the heterostructure slower, the switching also gets slower and less energy is dissipated in the switching process, elucidating the delay-energy trade-off for the device. The results show that switching in sub-nanosecond delay is plausible while dissipating energy of only $\sim$1 aJ. The ``$CV^2$'' energy dissipation is a couple of orders of magnitude lower than the energy dissipation due to Gilbert damping and it decreases with the increase of ramp period since the switching becomes more adiabatic. The standard deviation in switching delay for ramp period of 0.1 ns is about 22 ps and it increases about twice when the ramp period is increased to 1 ns. At higher ramp period, thermal fluctuations have more time to scuttle the magnetization and cause variability in switching time. Simulations have been performed at an elevated temperature (400 K) and the performance metrics switching delay and energy dissipation turn out to be similar (within 5\%) compared to that of room-temperature (300 K) case~\cite{supplx_add_sim}.

\section{Discussions}
\label{sec:discussions}

The giant magnetoresistance (GMR)~\cite{RefWorks:433,RefWorks:434} of the trilayer in this interface-coupled structure is calculated to be of the order of 30\%~\cite{RefWorks:649,RefWorks:688}, which provides an way to read the magnetization states (P-alignment or AP-alignment). Although this GMR is not that high compared to tunneling magnetoresistance (TMR)~\cite{RefWorks:555,RefWorks:572,RefWorks:76,RefWorks:74,RefWorks:33}, suitable design strategies can be possibly be devised to work with this moderate value of GMR and also it may be possible to increase the GMR by suitable material choice and design. It is also argued that even with the variance in the smaller thicknesses of the layers, it is still possible to couple the polarization and magnetization interracially in the proposed structures~\cite{RefWorks:649}. The modeling of interface anisotropy is not limited to the way we performed here, however, any \emph{strong} interface-coupled system would facilitate switching of magnetization from one state to the another. 

Due to the small lateral dimensions of these interface-coupled multiferroic heterostructures, it is possible to cram an enormous amount of devices on a single chip. Using an area density of $10^{-12}$ cm$^{-2}$, the dissipated power would be 10 mW/cm$^2$ considering 1 aJ energy dissipation in a single nanomagnet with 1 ns switching delay and 10\% switching activity (i.e., 10\% of the magnets switch at a given time). Such extremely dense and ultra-low-energy non-volatile computing systems can be powered by energy harvesting systems~\cite{roundy03,anton,lu,jeon}. 

\section{Conclusions}
\label{sec:conclusions}

In conclusion, we have performed an analysis over the switching dynamics of magnetization in interface-coupled multiferroic magnetoelectrics. We have modeled the interface anisotropy and calculated the performance metrics e.g., switching delay and energy dissipation as a function ramp period to elucidate the delay-energy trade-off. The results show that switching can take place in sub-nanosecond switching delay while expending $\sim$1 aJ of energy. Also, strong interface coupling facilitates error-resiliency during the switching and allows to scale down the lateral area to very small dimensions. Due to these superior performance characteristics of interface-coupled multiferroics, it would be of immense interest to work out different possible theoretical designs and experimental implementations. Successful experimental implementations must tackle the issue of process variation at low dimensions. Processors built on such platform can harbinger unprecedented applications that can work by harvesting energy from the environment e.g., medically implanted device to warn an impending epileptic seizure by monitoring the brain signals, wearable computers powered by body movements etc.

\section*{References}
\providecommand{\newblock}{}

\end{document}


\title[]{Supplementary Information\\Electric field-induced magnetization switching in interface-coupled multiferroic heterostructures: A highly-dense, non-volatile, and ultra-low-energy computing paradigm}

\author{Kuntal Roy}
\address{School of Electrical and Computer Engineering, Purdue University, West Lafayette, IN 47907, USA.\\**Some works were performed prior joining at Purdue University.}

\ead{royk@purdue.edu}

\date{\today}

\maketitle

\section{Switching delay-energy at T = 400 K}
\label{sec:S1}

\begin{figure}[h]
\centering
\includegraphics[width=80mm]{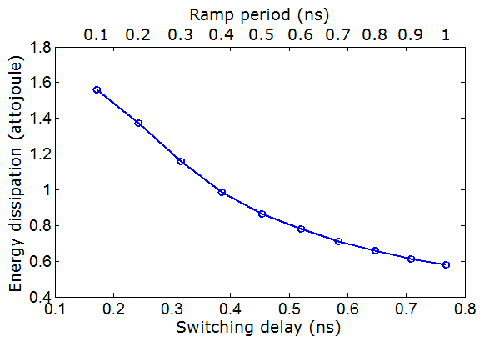}
\caption{\label{fig:delay_energy_ramp_400K} Switching delay-energy trade-off as a function of ramp period (upper axis) at T = 400 K. For a faster ramp, the switching becomes faster but the energy dissipation goes higher. Each point is generated from 10000 simulations in the presence of thermal fluctuations (400 K) and the average values of switching delays and energy dissipations are plotted. For 0.1 ns ramp period, the mean (max) switching delay is 172.3 ps (325 ps), while the mean energy dissipation is 1.56 aJ. For a slower ramp with period 1 ns, the mean (max) switching delay is 768 ps (970.2 ps), while the mean energy dissipation is 0.58 aJ.} 
\end{figure}

To get an understanding on the performance metrics switching delay and energy at elevated temperatures, we have performed simulations at T = 400 K, which is particularly important for burn-in testing of the devices. Fig.~\ref{fig:delay_energy_ramp_400K} shows the plot for switching delay versus energy dissipation at T = 400 K for different ramp periods (0.1 ns -- 1 ns). A moderately large number of simulations (10000) in the presence of thermal fluctuations (T = 400 K) have been performed to generate each point in the curve. When the magnetization reaches $\theta \leq 5^\circ$, the switching is deemed to have completed. With the increase in switching delay, less energy is dissipated in the switching process, elucidating the delay-energy trade-off for the device. 

\section{Comparison of performance metrics for T = 300 K and T = 400 K}
\label{sec:S2}

\begin{figure}
\centering
\includegraphics[width=80mm]{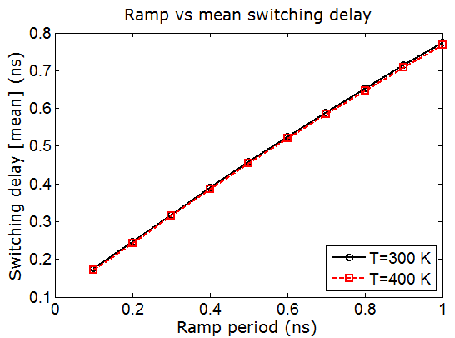}
\caption{\label{fig:ramp_mean_delay_temp_comparison} Mean switching delay versus ramp period for T=300 K and T=400 K. Each point is generated from 10000 simulations in the presence of thermal fluctuations. At the higher temperature of T=400 K, the mean switching delay decreases compared to that of T=300 K. For 0.1 ns ramp period, the mean switching delay at T = 300 K (400 K) is 175.3 ps (172.3 ps). For a slower ramp with period 1 ns, the mean switching delay at T = 300 K (400 K) is 775.2 ps (768 ps).} 
\end{figure}

We perform a comparison of switching delay (both mean and standard deviation) for the cases T = 300 K and T = 400 K. Figure~\ref{fig:ramp_mean_delay_temp_comparison} depicts that the mean switching delay at a higher temperature T = 400 K decreases compared to the case at T = 300 K. This can be conceived by the reasoning that the initial deflection of magnetization due to thermal fluctuations increases at a higher temperature. Hence, magnetization is likely to start more far from the easy axis at a higher temperature for different trajectories, leading to the decrease in switching delay. However, this decrease in mean switching delay at T = 400 K is very small (less than 2\%) compared to that of T = 300 K.

\begin{figure}
\centering
\includegraphics[width=80mm]{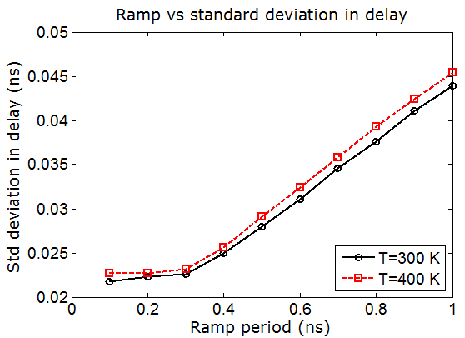}
\caption{\label{fig:ramp_std_deviation_delay_temp_comparison} Standard deviation in switching delay versus ramp period for T=300 K and T=400 K. Each point is generated from 10000 simulations in the presence of thermal fluctuations. At the higher temperature of T=400 K, the standard deviation in switching delay increases compared to that of T=300 K. For 0.1 ns ramp period, the standard deviation in switching delay at T = 300 K (400 K) is 21.8 ps (22.7 ps). For a slower ramp with period 1 ns, the standard deviation in switching delay at T = 300 K (400 K) is 43.9 ps (45.4 ps).} 
\end{figure}

Figure~\ref{fig:ramp_std_deviation_delay_temp_comparison} shows the standard deviation in switching delay for temperatures 300 K and 400 K. The trend of standard deviation in switching delay with the increase in temperature shows an opposite trend to that of the mean. The standard deviation at higher temperature increases with temperature, which bodes well with the Eq.~(3) in the main paper and also physically conceivable. This increase in standard deviation in switching delay at T = 400 K is quite small (less than 5\%) compared to that of T = 300 K.

The mean energy dissipation decreases with increasing temperature and this decrease at T = 400 K is quite small (less than 3\%) compared to the case of T = 300 K.

